\documentclass[runningheads]{llncs}
\usepackage[T1]{fontenc}
\usepackage{graphicx}
\usepackage{amsmath,amsfonts}
\usepackage{algorithmic}
\usepackage{algorithm}
\usepackage{array}
\usepackage{textcomp}
\usepackage{stfloats}
\usepackage{url}
\usepackage{verbatim}
\usepackage{graphicx}
\usepackage{cite}
\usepackage{multirow} 
\usepackage{float}
\usepackage{colortbl}
\graphicspath{ {image/} }
\usepackage{vcell}
\usepackage{adjustbox}
\usepackage{url}
\usepackage{makecell}
\usepackage{multirow}
\usepackage{rotating}
\usepackage{threeparttable} 
\usepackage{amsmath}  
\usepackage{amsfonts,amssymb}

\usepackage{booktabs}
\usepackage{arydshln}
\usepackage[misc]{ifsym}
\usepackage{xcolor}
\usepackage{cite}

\usepackage{xspace}
\newcommand{\etal}{\emph{et al.}\xspace}

\newcommand{\ie}{\emph{i.e.,}\xspace}
\newcommand{\etc}{\emph{etc.}\xspace}

\begin{document}
\title{Facilitating Feature and Topology Lightweighting: An Ethereum Transaction Graph Compression Method for Malicious Account Detection}
\titlerunning{Facilitating Feature and Topology Lightweighting}
%
\author{
Jiajun Zhou\inst{1,2} \and
Xuanze Chen\inst{1,2} \and
Shengbo Gong\inst{1,2} \and
Chenkai Hu\inst{3} \and  \\
Chengxiang Jin\inst{1,2} \and 
Shanqing Yu\inst{1,2} \and
Qi Xuan\inst{1,2} \textsuperscript{(\Letter)}
}

\authorrunning{Zhou \etal}
%
\institute{
Institute of Cyberspace Security, Zhejiang University of Technology, \\Hangzhou 310023, China  
\and
Binjiang Institute of Artificial Intelligence, ZJUT, \\Hangzhou 310056, China\\
\and
Polytechnic Institute, Zhejiang University, \\Hangzhou 310015, China\\
\email{xuanqi@zjut.edu.cn}}
%
\maketitle              
\begin{abstract}

Ethereum has become one of the primary global platforms for cryptocurrency, playing an important role in promoting the diversification of the financial ecosystem. However, the relative lag in regulation has led to a proliferation of malicious activities in Ethereum, posing a serious threat to fund security. Existing regulatory methods usually detect malicious accounts through feature engineering or large-scale transaction graph mining. However, due to the immense scale of transaction data and malicious attacks, these methods suffer from inefficiency and low robustness during data processing and anomaly detection. In this regard, we propose an \textbf{Eth}ereum \textbf{T}ransaction \textbf{G}raph \textbf{C}ompression method named TGC4Eth, which assists malicious account detection by lightweighting both features and topology of the transaction graph. At the feature level, we select transaction features based on their low importance to improve the robustness of the subsequent detection models against feature evasion attacks; at the topology level, we employ focusing and coarsening processes to compress the structure of the transaction graph, thereby improving both data processing and inference efficiency of detection models. Extensive experiments demonstrate that TGC4Eth significantly improves the computational efficiency of existing detection models while preserving the connectivity of the transaction graph. Furthermore, TGC4Eth enables existing detection models to maintain stable performance and exhibit high robustness against feature evasion attacks.

\end{abstract}

\section{Introduction}
The fintech sector is currently witnessing significant attention towards blockchain technology, driven by its attributes of anonymity, decentralization, and immutability. These attributes have captured the interest of a vast user base and propelled the growth of cryptocurrency transaction. 
Ethereum, being one of the most influential blockchain platforms, facilitates the creation and deployment of smart contracts, thereby further amplifying its influence in both financial and non-financial sectors.
Recently, the widespread application of cryptocurrencies in the financial sector has also brought new security challenges, leading to malicious activities such as phishing attacks, Ponzi schemes, ICO frauds, and contract vulnerability manipulation. In the struggle against regulatory technologies, these malicious behaviors continue to evolve, with increasingly varied forms and enhanced concealment. Therefore, effectively identifying and monitoring malicious accounts, accurately detecting and promptly addressing these issues, are crucial not only for protecting user assets but also for maintaining the stable development of the blockchain ecosystem.

Existing methods primarily concentrate on manual feature engineering or transaction graph mining, combined with machine learning techniques, to detect malicious accounts. However, these methods have some limitations. On the one hand, manual feature engineering captures the behavioral patterns of malicious accounts through elaborate multi-dimensional features, which not only relies on expert knowledge, but is also easily circumvented by new malicious patterns. On the other hand, given the vast amount of transaction data, graph mining methods often rely on sampling techniques to balance the scale of data, which can compromise the integrity of transaction information.


To solve these issues, we propose a \textbf{T}ransaction \textbf{G}raph \textbf{C}ompression method for \textbf{Eth}ereum malicious account detection, named TGC4Eth, which lightweights the transaction graph at both the feature and topology levels.
At the feature level, we first construct multi-dimensional transaction features and rank these features by importance, selecting those are difficult for attackers to evade for downstream detection tasks. At the topology level, we first capture the most relevant transaction subgraphs to the target nodes by graph focusing, and then perform graph coarsening to reduce the size of the transaction graph. By performing dual compression on the transaction graph, we maintain the connectivity of the graph while enhancing the concealment of transaction features during data processing.
We conduct extensive experiments on Ethereum transaction data, which demonstrates that our TGC4Eth can improve the computational efficiency and robustness of existing detection models in malicious account detection, while simultaneously preserving the performance stability of these models.


\section{Related Work}
\label{related work}
\subsection{Traditional Detection Methods}
Malicious account detection primarily concentrate on contract code~\cite{shen2021mining} and transaction records~\cite{lin2023ethereum,chen2019exploiting,farrugia2020detection,zhou2022behavior}. The former mainly analyzes the logic of smart contracts to determine whether they contain malicious risks, offering the advantage of early identification. The latter utilizes vast transaction data to analyze behavioral patterns of accounts and thereby assess whether they are involved in malicious activities. This approach is relatively flexible but exhibits a lag compared to the former.


Early studies mainly detect malicious accounts through feature engineering combined with machine learning methods.
Farrugia \etal~\cite{farrugia2020detection} designed 42 transaction-related features to detect multiple malicious behaviors simultaneously and analyzed the contribution of different features to the detection model.
Building on this, Ibrahim \etal~\cite{ibrahim2021illicit} refined the number of features used for detection to six through correlation analysis. 
Luo \etal~\cite{luo2023ai} employed statistical models, natural language processing techniques, and other machine learning models to detect emerging fraud related to DeFi (Decentralized Finance). 
Furthermore, some studies combine transaction features with contract code features for malicious account detection. 
Chen \etal~\cite{chen2018detecting} extracted account features and opcode features, applying ensemble learning methods such as Random Forest and XGBoost to identify Ponzi schemes.
Zhang \etal~\cite{zhang2021detecting} utilized account features, opcode features, and bytecode features, and enhanced the LightGBM~\cite{ke2017lightgbm} to detect Ponzi contracts. 
Galletta \etal~\cite{galletta2023sharpening} combined labeled data from the above three works and merged them to get 673 Ponzi accounts and analyzed the significance of 28 statistical features using the interface provided by the Etherscan website.

\subsection{Graph-based Detection Methods}
The design of manual features relies heavily on expert knowledge and the detection effectiveness is limited by their expressive power. As a result, several studies have exploited the properties of blockchain data to construct code graphs or large-scale transaction graphs and combine them with graph intelligence algorithms to automatically mine account behavioral features and perform malicious account detection.
Specifically, using traditional software engineering techniques, contract codes can be transformed into data flow graphs, control flow graphs~\cite{zhu2021similarity,zhuang2021smart}, abstract syntax trees~\cite{cai2024fine}, \etc, forming graph-structured data with variables and functions as nodes, and control or data flows as edges. And transaction records can be naturally constructed as transaction graphs with accounts as nodes and transactions as edges.
Bartoletti \etal~\cite{bartoletti2020dissecting} categorized Ponzi schemes into four transaction patterns: tree, chain, waterfall, and privilege transfer. 
Liang \etal~\cite{liang2024ponziguard} pioneered constructing a contract execution behavior graph, where transactions trigger variables forming edges. 
Chen \etal~\cite{chen2020phishing} extracted transaction subgraphs and used graph autoencoders to learn account features, ultimately identifying phishing accounts through LightGBM. 
Shen \etal~\cite{shen2021identity} proposed an end-to-end GCN model based on transaction subgraphs to detect phishing accounts in Ethereum and bot accounts in EOSIO. 
Jin \etal~\cite{jin2022dual} considered both code features and transaction features, building a dual-channel framework for Ponzi scheme alerts. 
Zhou \etal~\cite{zhou2022behavior} proposed a de-anonymization model that integrates hierarchical attention and contrast mechanisms, effectively learning node-level account features and subgraph-level account behavior patterns, aiding in identifying account identity types. 
Jin \etal~\cite{jin2022heterogeneous,jin2024time} constructed Ethereum transaction records into heterogeneous interaction graphs and designed static and temporal meta-paths to capture account behavior patterns.

\begin{figure}[t]
    \centering\includegraphics[width=0.9\textwidth]{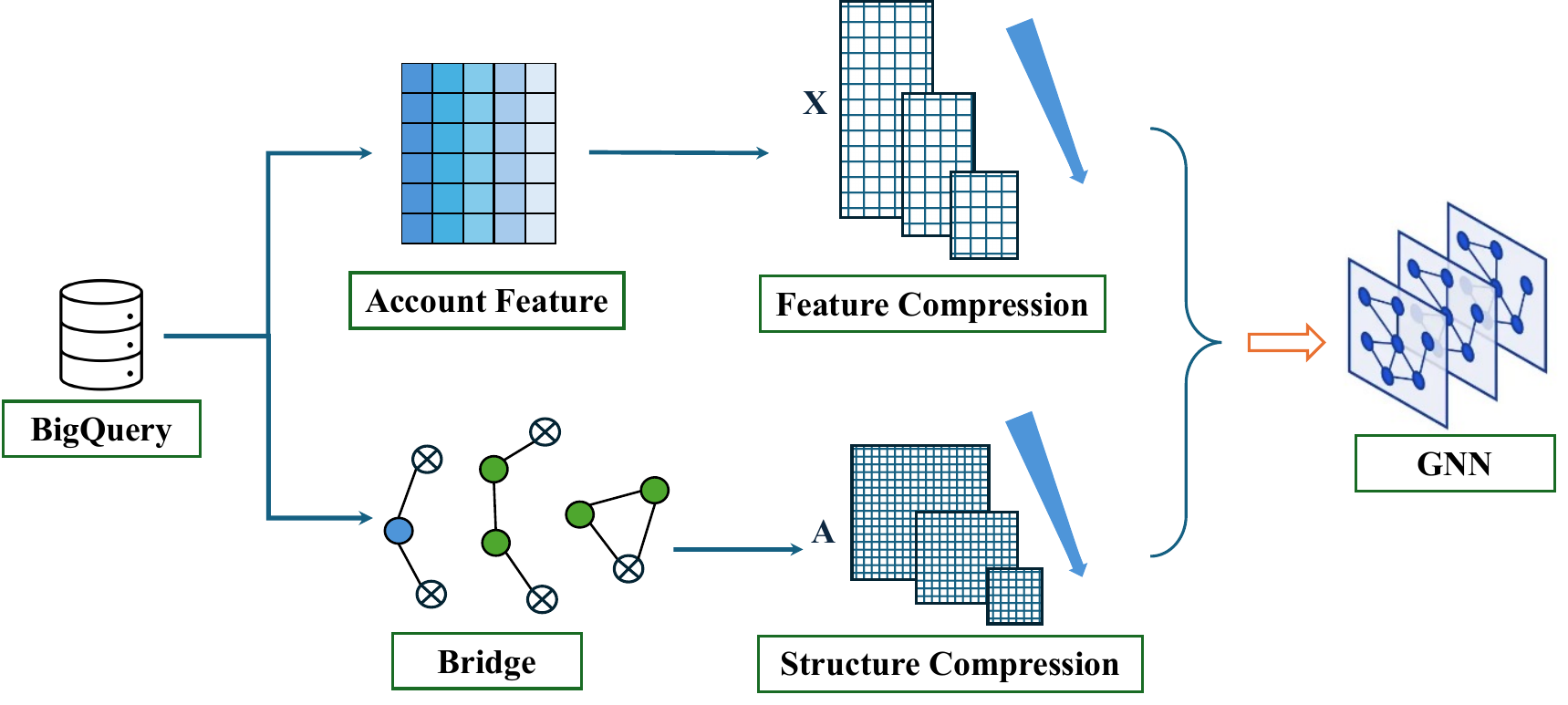}
    \caption{Overall framework of TGC4Eth.}
    \label{fig1}
\end{figure}

\section{Method}
\label{method}
In this section, we introduce our transaction graph compression method, as illustrated in Fig.~\ref{fig1}, which assists malicious account detection by lightweighting both features and topology of the transaction graph.

\subsection{Transaction Feature Compression}
\subsubsection{Feature Construction}
Our study period spans from 2018-01-01 to 2020-01-01. During feature extraction, we select the intersection of the top 10 important features derived from the three studies~\cite{farrugia2020detection,zhou2022behavior,galletta2023sharpening}. 
It is worth noting that while these studies listed all features, they did not systematically categorize them from multiple perspectives. Therefore, we reclassify these features and conduct a comprehensive analysis of each. 
Additionally, we introduce several new statistics as transaction features. All monetary values in this paper are denominated in Ethereum (ETH). 
Lifecycle refers to the time span (in minutes) from the first transaction to the last transaction of an account within the study window.
Different from previous studies, we count transaction frequency within the lifecycle rather than across the entire study period.


\begin{table}[t]
\renewcommand\arraystretch{1.3}      
\centering
\arrayrulecolor{black}
\caption{Summary of manual transaction features.}
\resizebox{\textwidth}{!}{                                             
   \begin{tabular}{lll} 
\hline\hline
Feature Name~                                    & Description                                           & Type                                                                                  \\ 
\hline
\textit{starting\_balance\_eth}                  & Initial Balance                                       & \multirow{3}{*}{Balance-related}                                                                \\
\textit{final\_balance\_eth}                     & Ending Balance                                        &                                                                                                 \\
\textit{diff\_balance\_eth}                      & Difference within the Research Window                 &                                                                                                 \\ 
\hline
\textit{total\_received\_eth}                    & Total revenue                                         & \multirow{5}{*}{Income-related}                                                                 \\
\textit{max\_value\_received\_eth}               & Single Transaction Maximum Revenue                    &                                                                                                 \\
\textit{min\_value\_received\_eth}               & Single Transaction Minimum Revenue                    &                                                                                                 \\
\textit{avg\_value\_received\_eth}               & Single Transaction Average Revenue                    &                                                                                                 \\
\textit{std\_value\_received\_eth}               & Standard Deviation of Revenue per Transaction         &                                                                                                 \\ 
\hline
\textit{total\_sent\_eth}                        & Total Expenditure                                     & \multirow{5}{*}{Expenditure-related}                                                            \\
\textit{max\_value\_sent\_eth}                   & Single Transaction Maximum Expenditure                &                                                                                                 \\
\textit{min\_value\_sent\_eth}                   & Single Transaction Minimum Expenditure                &                                                                                                 \\
\textit{avg\_value\_sent\_eth}                   & Single Transaction Average Expenditure                &                                                                                                 \\
\textit{std\_value\_sent\_eth}                   & Standard Deviation of Expenditure per Transaction     &                                                                                                 \\ 
\hline
\textit{max\_single\_neighbor\_count}            & Maximum Number of Transactions from a Neighbor        & \multirow{4}{*}{\begin{tabular}[c]{@{}l@{}}Neighbor-related \\(Undirected)\end{tabular}}        \\
\textit{max\_single\_neighbor\_value\_eth}       & Total Amount of Transactions from a Neighbor          &                                                                                                 \\
\textit{avg\_single\_neighbor\_count}            & Average Number of Transactions from a Neighbor        &                                                                                                 \\
\textit{avg\_single\_neighbor\_value\_eth}       & Average Total Amount of Transactions from a Neighbor  &                                                                                                 \\ 
\hline
\textit{num\_received\_single\_neighbor}         & Number of Unique Payee Neighbors                      & \multirow{7}{*}{\begin{tabular}[c]{@{}l@{}}Neighbor-related\\~(Directed)\end{tabular}}          \\
\textit{num\_sent\_single\_neighbor}             & Number of Unique Payer Neighbors                      &                                                                                                 \\
\textit{diff\_rs\_neighbor\_count}               & Difference in the Number of Payee and Payer Neighbors &                                                                                                 \\
\multirow{2}{*}{\textit{std\_dev\_received}}     & Standard Deviation of the Number of Payments          &                                                                                                 \\
                                                 & between~Unique Payer Neighbors                        &                                                                                                 \\
\multirow{2}{*}{\textit{std\_dev\_sent}}         & Standard Deviation of the Number of Receipts          &                                                                                                 \\
                                                 & between Unique Payee Neighbors                        &                                                                                                 \\ 
\hline
\textit{lifecycle\_min}                          & Account Lifecycle                                     & \multirow{5}{*}{\begin{tabular}[c]{@{}l@{}}Lifecycle and \\Transaction Frequency\end{tabular}}  \\
\textit{avg\_min\_between\_sent\_tnx}            & Average Number of Expenditures per Minute             &                                                                                                 \\
\textit{avg\_min\_between\_sent\_value\_eth}     & Average Amount of Expenditure per Minute              &                                                                                                 \\
\textit{avg\_min\_between\_received\_tnx}        & Average Number of Incomes per Minute                  &                                                                                                 \\
\textit{avg\_min\_between\_received\_value\_eth} & Average Amount of Income per Minute                   &                                                                                                 \\ 
\hline
\textit{if\_sc}                                  & Is it a Smart Contract?                               & \multirow{2}{*}{Account Type}                                                                   \\
\textit{if\_token}                               & Is it a Token?                                        &                                                                                                 \\
\hline\hline
\end{tabular}
}
\arrayrulecolor{black}
\label{tab:features}
\end{table}

The manual features summarized in Table~\ref{tab:features} encompass a comprehensive range of aspects, including balance, transaction amount, transaction frequency, income and expenditure. These features are more extensive compared to previous studies. 
Specifically, we summary a total of 29 features from seven perspectives. Transaction balances reflect the initial and final outcomes of the account. Single transaction features enable the acquisition of fine-grained directed transaction characteristics. Life cycle-related features indicate the transaction frequency during the account's active period. Account-type features help bridge the gap between homogeneous and heterogeneous transaction graphs. Benefiting from the performance of BigQuery, the feature extraction method in this paper is capable of efficiently handling massive datasets and achieving real-time statistics.

\subsubsection{Feature Selection}
\begin{figure}[t]
  \centering
  \includegraphics[width=0.9\textwidth]{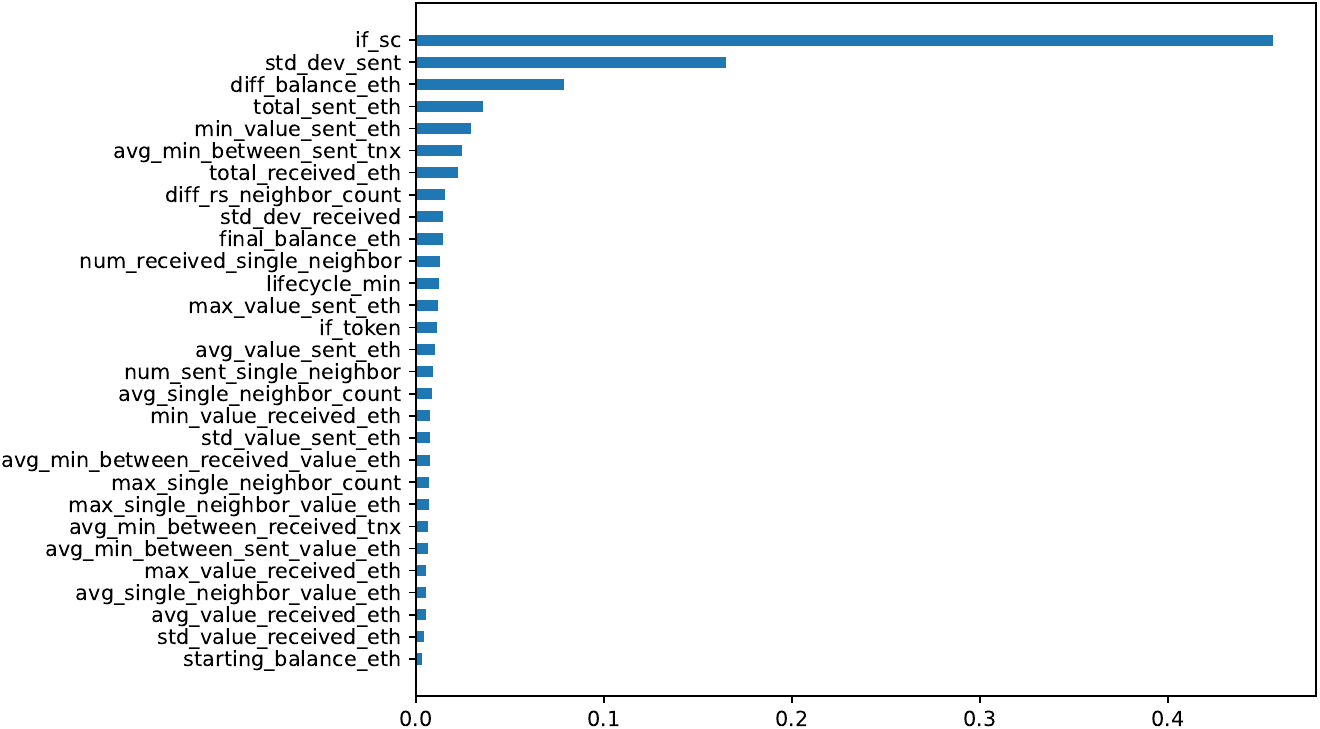}
  \caption{The importance of the manual transaction features derived by LightGBM.}
  \label{fig: importance}
\end{figure}

Due to the accessibility of machine learning-based detection models, adversaries might analyze the decision boundaries of these models and selectively adjust their behavioral patterns to weaken or even evade more significant features, thereby evading detection. To effectively counteract such potential feature evasion attacks, we adopt a low importance-based feature selection strategy that aims to compress transaction features to improve detection robustness. In this paper, we utilize the gain-based settings in LightGBM to quantify the importance of different features for detection, with the results illustrated in Fig~\ref{fig: importance}.
We conduct feature selection by evaluating feature importance and incorporating real-world analysis.
For example, the balance difference at the beginning and end of the time window is a commonly utilized metric that reflects the net income of the account during the studied period. 
Malicious accounts typically transfer assets after being flagged, resulting in a significant alteration of this metric. 
However, malicious accounts are not labeled in the actual detection, resulting in no significant difference in this feature.
Finally, We select 9 features out of the original 29 to characterize the attack based on their lower relevance to malicious behavior, including the following:
\begin{multicols}{2}
    \begin{itemize}
      \item[$\bullet$] \textit{starting\_balance\_eth}
      \item[$\bullet$] \textit{max\_value\_received\_eth}
      \item[$\bullet$] \textit{avg\_value\_received\_eth}
      \item[$\bullet$] \textit{std\_value\_received\_eth}
      \item[$\bullet$] \textit{max\_single\_neighbor\_count}
      \item[$\bullet$] \textit{max\_single\_neighbor\_value\_eth}
      \item[$\bullet$] \textit{avg\_single\_neighbor\_value\_eth}
      \item[$\bullet$] \textit{avg\_min\_between\_sent\_value\_eth}
      \item[$\bullet$] \textit{avg\_min\_between\_received\_tnx}
    \end{itemize}
\end{multicols}
Although removing highly important features might somewhat impact detection performance, relying on the feature robustness of GNN-based models allows us to maintain relatively stable detection capabilities. By employing feature compression, we can effectively counter potential feature evasion attacks, thereby enhancing the robustness of existing detection models.

\subsection{Transaction Topology Compression}

\begin{figure*}[t]
  \centering
  \includegraphics[width=0.8\textwidth]{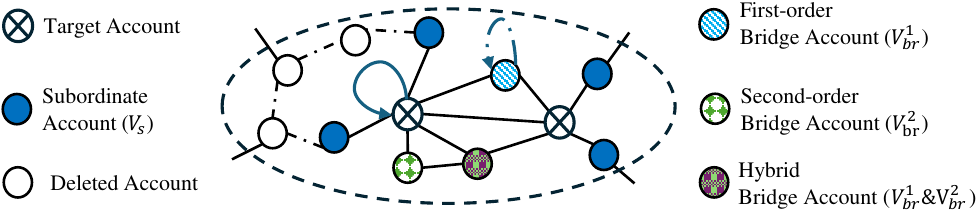}
  \caption{The illustration of initial transaction graph and different types of accounts.}
  \label{Fig: define}
\end{figure*}

After feature extraction, the information of multiple transactions between accounts has been embodied in these features, including transaction direction, amount, and frequency. Consequently, we only construct undirected transaction graphs.
Moreover, since all internal transactions are essentially various cascading transaction behaviors triggered by external transactions, we only consider external transactions in this paper.
The initial transaction graph is represented as $G_\textit{I} = (\mathcal{V}_t,\mathcal{V}_{ba},\mathcal{E})$, where $V_t\in\mathcal{V}_t$ represents the labeled accounts, referred to here as target accounts, while the others $V_\textit{ba}\in\mathcal{V}_\textit{ba}$ are called background accounts.
We then define bridge accounts $V_\textit{br}$ as those that exist on the paths connecting any two target accounts, denoted as 
$\{V_{t_i},V_{ba_1}, \cdots,V_{ba_n},V_{t_j}\}$.
The number of bridge accounts on the path is used to define the order of the bridge accounts. For instance, common neighbors of two target accounts can be considered as first-order bridge accounts $V_\textit{br}^1$. If a path includes two bridge accounts, then both of them are considered second-order bridge accounts $V_\textit{br}^2$. Note that some bridge accounts may be located on multiple paths between target accounts and are considered as hybrid bridge accounts.
Accounts that are first-order neighbors of the target account but are not bridge accounts are defined as subordinate accounts $V_\textit{s}$.
Fig.~\ref{Fig: define} illustrates the definition of initial transaction graph.

\begin{figure*}[t]
  \centering
  \includegraphics[width=0.9\textwidth]{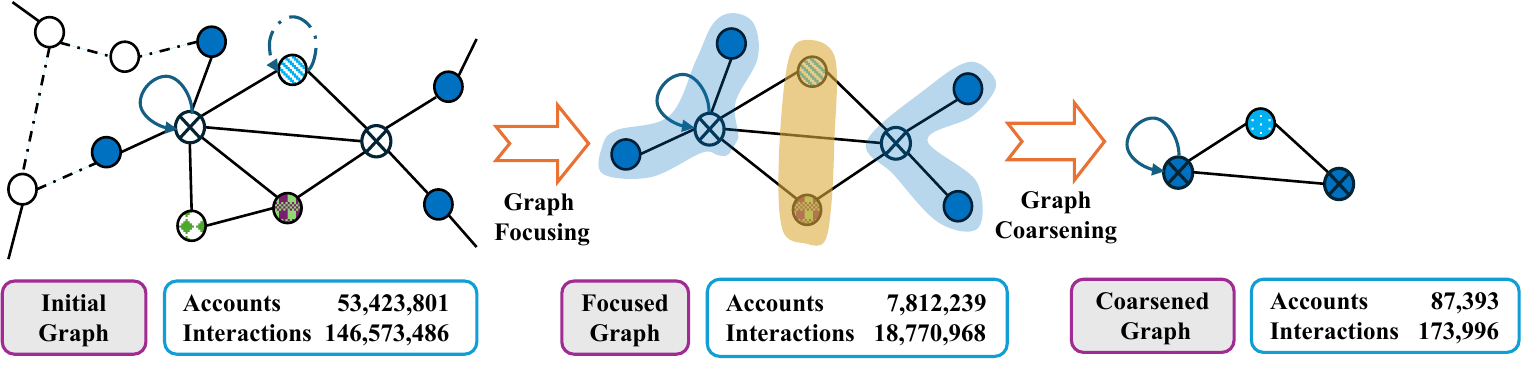}
  \caption{A framework for graph structure compression, including graph focusing and graph coarsening. The meaning of nodes and edges is the same as in Fig. 3.}
  \label{fig: topology}
\end{figure*}

\subsubsection{Graph Focusing}

During graph focusing, we first extract the first-order neighbors of the target accounts, then identify and label the bridge accounts and subordinate accounts among them. To compress the graph topology while ensuring graph connectivity, we retain first-order and second-order bridge accounts among the first-order neighbors. Furthermore, if there are hybrid bridge accounts on the path between any two target accounts, \ie these target accounts can be connected solely through first-order bridge accounts, we remove the redundant second-order bridge accounts. Ultimately, we obtain the focused graph $G_{F}$.

\subsubsection{Graph Coarsening}
The initial transaction graph contains approximately fifty million nodes. Although graph focusing has reduced this number to the millions, this scale remains computationally challenging for downstream detection models. Further analysis of the focused graph reveals that the importance of subordinate accounts is less than that of bridge accounts, but they are more numerous. To address this, we propose a graph coarsening method via account information aggregation to further compress the transaction graph topology. Specifically, we first aggregate the information of subordinate accounts into the target accounts, formalized as follows:
\begin{equation}
    \label{eq1}
    \hat{\boldsymbol{X}}_{t} =  \frac{1}{|\mathcal{N}_{s}+1|} \left( \boldsymbol{X}_{t} + \sum_{n \in \mathcal{N}_{s}}  \boldsymbol{X}_{s_n}   \right)
\end{equation}
where $\mathcal{N}_{s}$ represents the set of subordinate accounts for the target account $V_t$, and $\boldsymbol{X}_\ast$ is the feature vector of account $V_\ast$, $\hat{\boldsymbol{X}}_t$ is a composite feature that aggregates the features of the target account and all subordinate accounts.

In addition, since bridge accounts serves as the key connecting module, we aggregate the information from all bridge accounts between two target accounts to form a new composite bridge account, formalized as follows:
\begin{equation}
    \begin{aligned}
        \label{eq2}
        \hat{\boldsymbol{X}}_\textit{br}^1 &=  \frac{1}{|\mathcal{N}_\textit{br}^1|} \left( \sum_{n \in \mathcal{N}_\textit{br}^1}  \boldsymbol{X}_{\textit{br}_n}^1   \right) \\
        \hat{\boldsymbol{X}}_\textit{br}^{2-l} ,\ \hat{\boldsymbol{X}}_\textit{br}^{2-r} &=  \frac{1}{|2 \cdot \mathcal{N}_\textit{br}^2|}   \left( \sum_{n \in \mathcal{N}_\textit{br}^{2-l}} \boldsymbol{X}_{\textit{br}_n}^{2-l},  \sum_{n \in \mathcal{N}_\textit{br}^{2-r}} \boldsymbol{X}_{\textit{br}_n}^{2-r}   \right)
    \end{aligned}
\end{equation}
where $\mathcal{N}_\textit{br}^1$ represents the set of first-order bridge accounts between target accounts pair, $\mathcal{N}_\textit{br}^{2}$ represents the set of second-order accounts, $-l$ and $-r$ indicate the relative positions of these second-order bridge accounts in the path connetcing the target account pair.
Note that the aggregation of second-order bridge accounts between target account pairs will generate two composite second-order bridge accounts.
Ultimately, we obtain the coarsened graph $G_{C}$.
Fig.~\ref{fig: topology} illustrates the processing of transaction topology compression.



\subsection{Malicious Account Detection}
After Transaction graph compression, we obtain a more lightweight transaction graph. When conducting malicious account detection, we input the graph into a GNN-based detection model. During message aggregation and feature updating, we obtain the final account representation, formulated as follows:
\begin{equation}
    \label{eq3}
    \mathbf{\boldsymbol{x}}_{i}^{(k)}=\gamma^{(k)}\left(\mathbf{\boldsymbol{x}}_{i}^{(k-1)},\bigoplus_{j\in\mathcal{N}(i)}\phi^{(k)}\left(\mathbf{\boldsymbol{x}}_{i}^{(k-1)},\mathbf{\boldsymbol{x}}_{j}^{(k-1)}\right)\right)
\end{equation}
where $\phi$ and $\gamma$ represent differentiable functions such as multilayer perceptrons (MLPs), $\oplus$ denotes the integration function, such as summation, averaging, and maximum. 
For classification problems, a softmax function is often required for normalization to output the predicted probabilities $\hat{\boldsymbol{y}}_i$.
\begin{equation}     
\label{eq4}
\widehat{\boldsymbol{y}}_i =\text{Softmax}\left(\mathbf{\boldsymbol{x}}_i^{(k)}\right)=\frac{e^{\mathbf{\boldsymbol{x}}_i^{(k)}}}{\sum_je^{\mathbf{\boldsymbol{x}}_j^{(k)}}}
\end{equation}
We use cross-entropy loss function here.
\begin{equation}
    \mathcal{L}= -\sum_i  \boldsymbol{y}_i\log(\hat{\boldsymbol{y}}_i)
\end{equation}
where $\boldsymbol{y}$ is the one-hot vector of account label. 

\section{Experiment}
\label{experiments}


\begin{table}[t]
    \caption{Statistics for each type of account in different compression graphs.}
    \centering
    \setlength{\tabcolsep}{10pt}
    \begin{tabular}{ccc}
    \hline\hline
    \multirow{2}{*}{Account Role} & \multicolumn{2}{c}{Quantities}  \\
                                      & $G_F$    & $G_C$             \\ 
    \hline
    Target Account $V_t$                       & 5,880     & 5,880               \\
    Subordinate Account $V_s$                  & 2,349,274 & 0                   \\
    First-order Bridge  $V_{br}^1$         & 2,668,199 & 81,513              \\
    Second-order Bridge  $V_{br}^2$         & 3,998,701 & 730,202             \\
    \hline\hline
    \end{tabular}
    \label{tab: dataset}
\end{table}

\subsection{Dataset}
In this section, we primarily discuss the data preparation process. We obtain the labels for accounts from \textit{Etherscan}\footnote[1]{\url{https://cn.etherscan.com}} and \textit{Cryptoscam}\footnote[2]{\url{https://cryptoscamdb.org}}. 
Since our study period is set from ``2018-01-01'' to ``2020-01-01'', we exclude accounts that had no transactions during this period, resulting in 5,880 labeled accounts that had transactions. The final dataset scale is shown in Table~\ref{tab: dataset}. We category malicious accounts into four types: 2,163 Phish/Hack, 1,257 Scamming, 18 Exploit, and 4 Unsafe, totaling 3,442. Normal accounts include platform, protocol, exchange, specific projects and applications, totaling 2,438.
The total number of accounts and transactions acquired is 53,423,801 and 493,998,644, respectively.
Additionally, we present the number of various bridge accounts during graph topology compression, as shown in Table~\ref{tab: dataset}. 
During graph focusing, we reduce the number of accounts from 50 million to 8 million and further decrease it to 800,000 through graph coarsening. 
Compared to the initial transaction graph, our graph compression method reduces the total number of accounts to approximately 1\%.



\subsection{Experimental Setup}
Since our proposed TGC4Eth aims to assist in detecting malicious accounts by lightweighting the transaction graph, we combine it with various GNN-based detection models, including GCN~\cite{kipf2017semi}, SGC~\cite{wu2019simplifying}, SAGE~\cite{hamilton2017inductive}, APPNP~\cite{gasteiger2018predict}. 
Meanwhile, we compare GNN-based detection methods with machine learning-based methods, including LightGBM and MLP, to illustrate the superiority of the former in handling compressed data. 
We set the number of layers for all GNN-based methods to 2, hidden layer dimension to 64, output head dimension to 2, training epochs to 500, and early stopping rounds to 100. We adopt the AdamW optimizer~\cite{loshchilov2018decoupled} and GeLU activation function, with the dropout set to 0 and the learning rate set to 0.05. 
The relevant parameters for LightGBM include \textit{objective} is binary, \textit{metric} is auc, \textit{n\_estimators} is 100. 
The data is standardized using z-score normalization before being inputted. Model-specific parameters are set to default.
The dataset is divided into the training, validation and testing sets, with proportions of 60\%, 20\%, and 20\% respectively. We report the average Accuracy and AUC with 5 repeated experiments.

\begin{table}[t]
\renewcommand\arraystretch{1.2}      
\centering
\arrayrulecolor{black}
\caption{Detection performance under different compression settings.}
\resizebox{\textwidth}{!}{ 
\begin{tabular}{c|cc|cc|cc|cc}
\hline\hline
Features   & \multicolumn{4}{c|}{Feat-29}                                        & \multicolumn{4}{c}{Feat-9}                                          \\ 
\hline
Graph & \multicolumn{2}{c|}{$G_{F}$} & \multicolumn{2}{c|}{$G_{C}$} & \multicolumn{2}{c|}{$G_{F}$} & \multicolumn{2}{c}{$G_{C}$}  \\ 
\hline
Metrics    & ACC (\%)       & AUC~(\%)        & ACC~(\%)       & AUC~(\%)        & ACC~(\%)       & AUC~(\%)        & ACC~(\%)       & AUC~(\%)        \\ 
\hline
LightGBM   & ~~~92.79\scriptsize{$\pm$0.54}~~~ & ~~~92.68\scriptsize{$\pm$0.61}~~~  & ~~~92.79\scriptsize{$\pm$0.54}~~~ & ~~~92.68\scriptsize{$\pm$0.61}~~~  & ~~~89.01\scriptsize{$\pm$0.87}~~~ & ~~~88.92\scriptsize{$\pm$0.96}~~~  & ~~~89.01\scriptsize{$\pm$0.87}~~~ & ~~~88.92\scriptsize{$\pm$0.96}~~~  \\
MLP        & 88.64\scriptsize{$\pm$0.45} & 87.57\scriptsize{$\pm$0.64}  & 88.64\scriptsize{$\pm$0.45} & 87.57\scriptsize{$\pm$0.64}  & 86.21\scriptsize{$\pm$0.50} & 84.07\scriptsize{$\pm$0.31}  & 86.21\scriptsize{$\pm$0.50} & 84.07\scriptsize{$\pm$0.31}  \\ 
\hline
GCN        & 90.19\scriptsize{$\pm$1.16} & 89.61\scriptsize{$\pm$1.15}  & 89.68\scriptsize{$\pm$0.95} & 89.42\scriptsize{$\pm$1.06}  & 88.06\scriptsize{$\pm$0.68} & 86.59\scriptsize{$\pm$0.77}  & 88.04\scriptsize{$\pm$0.58} & 86.95\scriptsize{$\pm$0.72}  \\
SGC        & 90.24\scriptsize{$\pm$0.80} & 89.64\scriptsize{$\pm$0.65}  & 90.00\scriptsize{$\pm$0.84} & 89.30\scriptsize{$\pm$1.28}  & 88.33\scriptsize{$\pm$0.62} & 87.00\scriptsize{$\pm$0.62}  & 88.09\scriptsize{$\pm$0.53} & 87.06\scriptsize{$\pm$0.76}  \\
SAGE       & 89.74\scriptsize{$\pm$0.98} & 89.00\scriptsize{$\pm$1.01}  & 90.85\scriptsize{$\pm$0.57} & 90.28\scriptsize{$\pm$0.39}  & 88.30\scriptsize{$\pm$1.66} & 89.00\scriptsize{$\pm$1.01}  & 89.63\scriptsize{$\pm$0.38} & 88.75\scriptsize{$\pm$0.48}  \\
APPNP      & 87.04\scriptsize{$\pm$0.92} & 87.01\scriptsize{$\pm$0.68}  & 89.37\scriptsize{$\pm$0.72} & 88.25\scriptsize{$\pm$0.56}  & 80.75\scriptsize{$\pm$4.09} & 81.42\scriptsize{$\pm$4.08}  & 88.17\scriptsize{$\pm$0.44} & 87.39\scriptsize{$\pm$0.56}  \\
\hline\hline
\end{tabular}
}
\arrayrulecolor{black}
\label{tab: result}
\end{table}


\subsection{Evaluation of Graph Compression}
To validate the effectiveness of our graph compression method, we compare the detection performance of ML-based and GNN-based methods.
Specifically, we conduct experiments using the Feat-29 and Feat-9 feature sets, respectively, in combination with focused and coarsened graphs. The feature sets are used to initialize the initial transaction graph features.
The results are reported in Table~\ref{tab: result}, from which we can draw the following conclusions:
\begin{itemize}
	\item[$\bullet$] ML-based methods are more sensitive to changes in features. With feature compression, the performance of LightGBM on the coarsened graph shows a decrease of 4.07\% in ACC and 4.06\% in AUC, while MLP exhibits reductions of 2.74\% in ACC and 4.00\% in AUC. In contrast, GNN-based methods suffer less performance loss, showing better robustness under feature variation;
    \item[$\bullet$] Feature compression on coarsened graphs has a smaller impact on the performance of GNN-based methods compared to focused graphs, indicating that the complete graph compression process can maintain the stability of the detection models' performance. However, APPNP suffers significant performance loss during the feature compression on focused graphs, which is due to the fact that the low importance features as well as the corruption of the transaction graph topology seriously affect the message aggregation process of APPNP, so that its retained initial features and the currently aggregated features cannot be effectively fused, which further highlights the importance of the graph coarsening process.
    \item[$\bullet$] In transaction graph detection after feature compression, GNN-based models generally perform better on coarsened graphs than that on focused graphs, and in most cases, slightly outperform ML-based models. This indicates the robustness of GNN-based models to graph compression, suggesting that our TGC4Eth can maintain the performance stability of GNN-based models.
\end{itemize}

\subsection{Robustness Analysis under Feature Evasion Attack}
\begin{figure}
	\centering
	\includegraphics[width=0.7\textwidth]{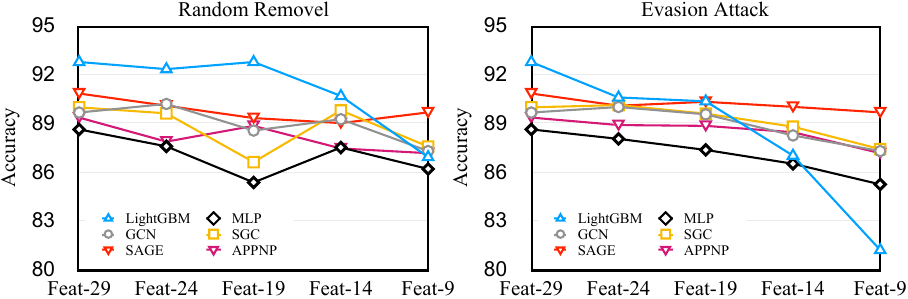}
	\caption{Performance under feature random removal and evading attack.}
	\label{fig: attack}
\end{figure}

To further evaluate the robustness of existing detection methods when faced with feature evasion attacks, we generate five different feature sets (Feat-29, Feat-24, Feat-19, Feat-14, Feat-9) via two feature compression methods: random removal and evasion attack. For a fair comparison, all the experiments are performed on the coarsened graph, and the results are shown in Fig.~\ref{fig: attack}.
\begin{itemize}
	\item[$\bullet$] We compare random removal and evasion attacks, using the latter to simulate an adversary's behavior in evading features. We find that evasion attacks targeting feature importance have a greater impact on detection performance than random removal, with all methods showing a decline in performance when facing evasion attacks, while results fluctuated under random removal. This is consistent with our expectations and further underscores the serious threat that feature evasion poses to detection performance.
    \item[$\bullet$] As the number of features decreased from 29 to 9, the ACC of GNN-based methods drop only by 1\%-2\%. This may be due to their semi-supervised neighborhood aggregation mechanism, which partially compensates for the loss of features using graph topology information. In contrast, ML-based models that utilize labeled data exhibit significant performance fluctuations, with LightGBM's ACC decreasing by as much as 6\% and 12\% under the two settings, respectively. This difference highlights the superior robustness of GNN-based methods compared to ML-based methods and further emphasizes that feature evasion attacks are a matter of concern.
\end{itemize}

\subsection{Graph Quality Analysis}
To assess the superiority of our graph coarsening method over random sampling in maintaining graph connectivity, we conduct a quality comparison analysis between the coarsened transaction graph $G_\textit{C}$ and a sampled transaction graph $G_\textit{R}$. We first define a connectivity metric that differs from the traditional one:
\begin{equation}
    \text{Connectivity}=\frac{\text{The number of nodes in the maximum connected component}}{\text{The number of node in the graph}}
\end{equation}
This metric assumes that a graph with better connectivity should have larger relative connected components. Table~\ref{tab: connect} shows the statistical differences between the coarsened and sampled graphs, from which it can be observed that, although random sampling yields more nodes, the number of edges is significantly lower, and the connectivity is also much less than that of the coarsened graph. This is because random sampling will generate many isolated communities, which are detrimental to message propagation. In contrast, our graph coarsening method effectively ensure graph connectivity while reducing the scale of transaction graph, which is beneficial for the training of subsequent detection models.
\begin{table}[t]
    \caption{Results of the comparison of graph structural integrity.}
    \centering
    \setlength{\tabcolsep}{10pt}
    \begin{tabular}{cccc} 
    \hline\hline
    Indicators     & $G_F$      & $G_R$  & $G_C$    \\ 
    \hline
    Accounts    & 7,812,239  & 94,654 & 87,393   \\
    Transactions    & 18,770,968 & 99,409 & 173,996  \\
    Average Degree & 4.8056     & 2.1004 & 3.9849   \\
    Connectivity   & 0.9999     & 0.3462 & 0.9862   \\
    \hline\hline
    \end{tabular}
    \label{tab: connect}
\end{table}

\section{Conclusion} \label{conclusion}
This paper presents a transaction graph compression method that effectively reduces data scale from both feature and topological perspectives. Experimental results demonstrate that our method can significantly enhance the computational efficiency of GNN-based detection methods. Additionally, this paper analyzes feature compression from the perspective of feature evasion attacks, confirming the robustness of GNN-based detection methods when faced with such attacks. However, this study also has some limitations, including the need for efficiency optimization in the graph compression process and the design of selection strategies during the feature compression process. Furthermore, the paper lacks downstream detection models tailored for the compressed graphs, thus it cannot guarantee that the graph compression method will yield optimal detection performance.

\subsubsection*{Acknowledgments.} 
This work was supported in part by the Key R\&D Program of Zhejiang under Grants 2022C01018 and 2024C01025, by the National Natural Science Foundation of China under Grants 62103374 and U21B2001.

\bibliographystyle{splncs04_}
\bibliography{sample-base}

\end{document}